\documentclass[review,12pt]{elsarticle}

\usepackage{lineno}

\usepackage{graphicx}
\usepackage{amsfonts}
\usepackage{amssymb}
\usepackage{float}
\usepackage{latexsym}
\usepackage{amsmath}

\usepackage{subfigure}
\usepackage[euler]{textgreek}
\usepackage{textcomp}
\usepackage{CJK}

\usepackage{algorithm}
\PassOptionsToPackage{noend}{algpseudocode}
\usepackage[end]{algpseudocode}

\usepackage[table]{xcolor}

\begin{document}

\begin{frontmatter}

\title{Information Transport and Observability in Compressible Aerodynamics}

\author[add1]{Bo Zhang\corref{cor1}}
\ead{bzhang@niu.edu}
\cortext[cor1]{Corresponding author. 
	Address:  
 	DeKalb, IL 60115, USA
 }
\address[add1]{Department of Mechanical Engineering, Northern Illinois University, DeKalb, IL 60115, USA}

\begin{abstract}
Pressure measurements provide sparse but direct observations of compressible aerodynamic flows, yet how information about hidden aerodynamic parameters is transported through the flow and encoded in these observations remains poorly understood. Here, we investigate information transport and observability in compressible aerodynamics using a differentiable shock-capturing immersed-boundary solver. By propagating gradients through the full unsteady flow solution, an automatic-differentiation-based observability metric is introduced to quantify the sensitivity of sparse pressure measurements to unknown aerodynamic parameters and identify informative sensing locations for inverse learning. The results reveal that aerodynamic information is transported non-uniformly through the flow field, producing localized regions of high observability. Inverse-learning experiments further demonstrate that observability and learnability are related but distinct concepts: although highly observable probes generally facilitate accurate parameter recovery, the highest-observability probe is not consistently the most effective for parameter inference. Furthermore, both the flow regime and the airfoil geometry substantially influence the distribution of observability and the convergence behavior of inverse learning. 
These findings establish a quantitative framework for understanding how aerodynamic information is encoded in sparse measurements and demonstrate the potential of automatic differentiation for observability analysis, informative sensor selection, and aerodynamic inverse analysis.
\end{abstract}


\end{frontmatter}



\section{Introduction}
Pressure measurements are widely used to infer aerodynamic conditions, monitor vehicle performance and support feedback control. 
Yet how information about hidden aerodynamic parameters is transported through a compressible flow and encoded in sparse observations remains poorly understood. 
While pressure signals are routinely used to estimate quantities such as angle of attack, Mach number and aerodynamic loads, the physical pathways through which this information becomes observable have not been systematically investigated. A fundamental question therefore arises: which regions of a compressible flow carry information about hidden aerodynamic parameters, and how can this information be exploited for reliable parameter inference?

Recent advances in flow-state estimation have shown that sparse pressure measurements can effectively constrain hidden flow states through data assimilation~\citep{Da_2020a}. Adjoint-variational approaches have reconstructed turbulent flows from highly limited measurements~\citep{Wang_2021b}, identified observable and unobservable structures in wall-bounded turbulence~\citep{Wang_2022b}, and have been extended to compressible and high-speed flows, including hypersonic boundary layers~\citep{Buchta_2022a} and shock-dominated systems~\citep{Ke_2026a}. In parallel, observable-augmented machine-learning approaches have incorporated physically meaningful observables into low-dimensional representations of turbulent flows, enabling improved analysis of multi-source flow data~\citep{Fukami_2025b}. Together, these studies demonstrate the growing potential of sparse measurements and physically informed observables for understanding and reconstructing complex flow phenomena. Here, however, we shift the focus from flow-state estimation to aerodynamic-parameter observability and investigate how information about hidden aerodynamic parameters is transported through compressible flows and encoded in sparse pressure measurements.

The problem is closely related to observability, sparse sensing and sensor placement. Observability-based approaches have been used to guide sensor placement in aerodynamic flows~\citep{Hinson_2014a}, while sparse sensing techniques have shown that high-dimensional systems can often be characterized using remarkably few measurements~\citep{Brunton_2016a, Loiseau_2018a, Fukami_2025a}. Information-theoretic and Gramian-based frameworks further relate measurement locations to parameter identifiability and estimation accuracy~\citep{Wang_2005a, Himpe_2018a}. Despite these developments, the observability of aerodynamic parameters in compressible flows, and its relationship to successful inverse learning, remains largely unexplored.

Compressible airfoil flows provide a particularly rich setting in which to address this question. Shock waves, expansion regions, shear layers, and wake structures shape the pressure field and influence how parameter-induced pressure disturbances are transported and encoded in sparse pressure measurements. Previous studies have demonstrated the important roles of shock-boundary-layer interactions, buffet oscillations, and large-scale flow structures in aerodynamic behavior~\citep{Hartmann_2013a, Li_2018a, Ma_2020a}. However, how these transport processes determine the observability of hidden aerodynamic parameters has not been quantified.

The present work investigates information transport and observability in compressible aerodynamics using a differentiable shock-capturing immersed-boundary solver~\citep{Zhang_2026a}. An automatic-differentiation-based observability metric is introduced to quantify the sensitivity of sparse pressure measurements to hidden aerodynamic parameters and identify informative sensing locations for inverse learning. Although the present paper focuses on pitching-amplitude inference, the proposed framework is applicable to a broad class of aerodynamic inverse problems, including the estimation of the angle of attack, Mach number, heaving motion, thermodynamic parameters, and coupled motions. Inverse-learning experiments are performed to examine the relationship between observability and parameter learnability under different sensor configurations, flow regimes, and airfoil geometries. The results reveal that aerodynamic information is encoded non-uniformly in sparse pressure measurements, producing localized regions of high observability whose effectiveness for parameter inference depends on both the flow regime and the underlying flow structures. These findings establish a physics-based framework for investigating information transport, observability, and inverse learnability in compressible aerodynamics, while demonstrating the potential of automatic differentiation for observability analysis, informative sensor selection, and aerodynamic inverse analysis.

\section{Differentiable Framework} 
\label{sec:approach}

\begin{figure}
\begin{center}
\includegraphics [width=1.\columnwidth]{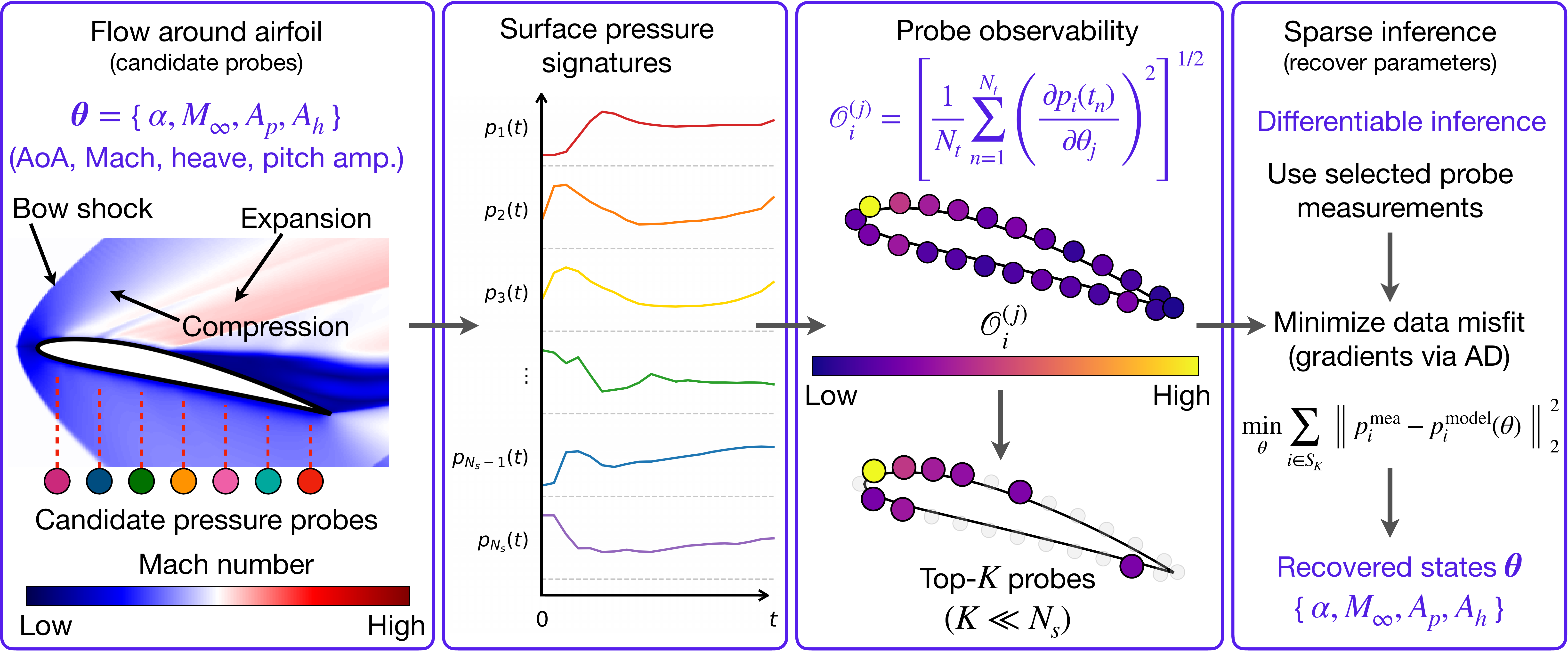}
\end{center}
\caption{Schematic of the differentiable observability framework.}
\label{fig:framework}
\end{figure}  

The objective of the present framework is to infer hidden aerodynamic parameters from sparse pressure measurements and to quantify how information about these parameters is transported through compressible flows and encoded in sparse observations. Figure~\ref{fig:framework} illustrates the overall methodology. Hidden aerodynamic parameters, including the angle of attack, Mach number, pitching amplitude and heaving amplitude, are not directly observable. Instead, they modify the compressible flow dynamics and leave distinct aerodynamic footprints within the flow field. Variations in these parameters alter shock locations, pressure distributions, expansion regions, shear layers and wake structures, which are subsequently reflected in the pressure signals recorded by sparse probes distributed around the airfoil and within the wake.

The central hypothesis of the present work is that hidden aerodynamic parameters generate observable pressure perturbations whose transport through the compressible flow determines their accessibility to sparse measurements. Although these parameters are not measured directly, their influence is redistributed by the evolving flow field and ultimately encoded in sparse pressure measurements. The inverse problem may therefore be interpreted as recovering hidden aerodynamic parameters from these observable aerodynamic footprints. This perspective unifies information transport, observability, and parameter inference within a differentiable framework.

In the present framework, sparse pressure measurements provide the available observations, while the governing equations describe the propagation of parameter-induced perturbations through the compressible flow. The flow solver is formulated as a differentiable dynamical system, enabling the sensitivities of sparse pressure measurements to hidden aerodynamic parameters to be computed directly through automatic differentiation. These sensitivities quantify how aerodynamic information is transported through the flow and encoded in sparse observations.

The conservative variables are defined as
\begin{equation}
\mathbf{U}
=
\left(
\rho,
\rho u,
\rho v,
E
\right)^T ,
\end{equation}
where $\rho$ denotes the density, $u$ and $v$ the velocity components, and $E$ the total energy.

The flow is governed by the two-dimensional compressible Navier--Stokes equations in conservative form,
\begin{equation}
\frac{\partial \mathbf{U}}{\partial t}
+
\frac{\partial \mathbf{F}}{\partial x}
+
\frac{\partial \mathbf{G}}{\partial y}
=
\mathbf{S}_{AV},
\label{eq:governing}
\end{equation}
where $\mathbf{F}$ and $\mathbf{G}$ denote the total flux vectors in the streamwise and transverse directions, each consisting of inviscid and viscous contributions. The source term $\mathbf{S}_{AV}$ represents the artificial-viscosity contribution introduced to stabilize shocks, discontinuities, and under-resolved gradients. Solid boundaries are treated using a ghost-cell immersed-boundary method~\citep{Mittal_2005a}, which is enforced through ghost-cell reconstruction and therefore does not appear as an explicit forcing term in the governing equations. 

Following spatial discretization, Eq.~(\ref{eq:governing}) can be written as the semi-discrete system
\begin{equation}
\frac{d\mathbf{U}}{dt}
=
\mathcal{R}(\mathbf{U},\boldsymbol{\theta}),
\label{eq:semi_discrete}
\end{equation}
where $\mathcal{R}$ denotes the differentiable residual operator and
\begin{equation}
\boldsymbol{\theta}
=
\left(
\alpha,\,
M_{\infty},\,
A_p,\,
A_h
\right)^{T},
\label{eq:parameters}
\end{equation}
is the hidden aerodynamic parameter vector, with $\alpha$ the angle of attack, $M_{\infty}$ the freestream Mach number, $A_p$ the pitching amplitude, and $A_h$ the heaving amplitude. The oscillation frequency and phase are prescribed and are therefore excluded from the inferred parameters. Although the formulation is presented for a general aerodynamic parameter vector, the numerical examples in the present study consider representative single-parameter inverse problems to isolate the relationship between sensor observability and inverse learnability. The residual operator incorporates the differentiable finite-volume discretization, high-order nonlinear reconstruction, approximate Riemann solver, immersed-boundary treatment, artificial-viscosity stabilization, freestream boundary conditions, and prescribed airfoil motion, enabling end-to-end differentiation through the complete flow solver via automatic differentiation.

The flow evolution is advanced in time according to
\begin{equation}
\mathbf{U}^{n+1}
=
\mathbf{U}^{n}
+
\Delta t\,
\mathcal{R}
\left(
\mathbf{U}^{n},
\boldsymbol{\theta}
\right),
\label{eq:time_update}
\end{equation}
where $\Delta t$ is the time step.
Repeated time integration defines the flow-evolution operator
\begin{equation}
\mathbf{U}^{N}
=
\Phi
\left(
\mathbf{U}^{0},
\boldsymbol{\theta}
\right),
\label{eq:solution_operator}
\end{equation}
which maps the initial flow state and hidden aerodynamic parameters to the flow field at the final time. Here, $\Phi$ denotes the complete time-marching solver from the initial condition to the final time. Different parameter vectors produce distinct parameter-dependent flow fields and pressure distributions, which provide the observable aerodynamic footprints used for inverse learning.

Sparse pressure measurements are extracted from the numerical solution through an observation operator
\begin{equation}
\mathbf{y}^{n}
=
\mathcal{H}
\left(
\mathbf{U}^{n}
\right),
\label{eq:observation}
\end{equation}
where $\mathbf{y}^{n}$ denotes the predicted pressure measurements at the sensor locations and $\mathcal{H}$ maps the flow field to the measurement space. The probes may be located on or near the airfoil surface, within the surrounding flow field, or in the downstream wake, thereby sampling the parameter-induced pressure perturbations transported through the flow.

The composition of the flow-evolution and observation operators defines the parameter-to-observation map
\begin{equation}
\mathcal{M}
=
\mathcal{H}
\circ
\Phi .
\label{eq:param_obs_map}
\end{equation}
This mapping describes how hidden aerodynamic parameters are transformed into observable pressure measurements through the governing equations. The inverse problem is therefore formulated as identifying the hidden aerodynamic parameters that generated the observed aerodynamic footprint.

Given the reference observations $\mathbf{y}^{n}_{\mathrm{ref}}$, the hidden aerodynamic parameters are inferred by minimizing the discrepancy between the predicted and reference pressure measurements. The loss function is defined as
\begin{equation}
\mathcal{L}(\boldsymbol{\theta})
=
\frac{1}{N_t N_s}
\sum_{n=1}^{N_t}
\left\|
\frac{
\mathbf{y}^n(\boldsymbol{\theta})
-
\mathbf{y}^n_{\mathrm{ref}}
}
{\rho_\infty u_\infty^2}
\right\|_2^2 ,
\label{eq:loss}
\end{equation}
where $N_t$ and $N_s$ are the numbers of sampled time instants and sensors, respectively, and $\|\cdot\|_2$ denotes the Euclidean norm. The pressure measurements at time $t_n$ are assembled into the observation vectors
\begin{equation}
\mathbf{y}^n
=
\begin{bmatrix}
p_1(t_n) \\
p_2(t_n) \\
\vdots \\
p_{N_s}(t_n)
\end{bmatrix},
\qquad
\mathbf{y}^{n}_{\mathrm{ref}}
=
\begin{bmatrix}
p_{1,\mathrm{ref}}(t_n) \\
p_{2,\mathrm{ref}}(t_n) \\
\vdots \\
p_{N_s,\mathrm{ref}}(t_n)
\end{bmatrix},
\label{eq:obs_vector}
\end{equation}
where $p_i(t_n)$ denotes the predicted pressure at the $i$th sensor and time $t_n$, and $p_{i,\mathrm{ref}}(t_n)$ is the corresponding reference pressure. The pressure mismatch is normalized by the freestream dynamic-pressure scale, $\rho_\infty u_\infty^2$, yielding a dimensionless mean-squared error over all sensor--time observations.

Because the entire flow solver is differentiable, the gradient of the loss function with respect to the hidden aerodynamic parameters is obtained through automatic differentiation (AD),
\begin{equation}
\nabla_{\boldsymbol{\theta}}
\mathcal{L}
=
\frac{\partial \mathcal{L}}
{\partial \boldsymbol{\theta}} .
\label{eq:gradient}
\end{equation}
These gradients quantify how perturbations of the hidden aerodynamic parameters propagate through the nonlinear compressible flow and influence the pressure measurements.
The parameters are updated iteratively according to
\begin{equation}
\boldsymbol{\theta}^{k+1}
=
\boldsymbol{\theta}^{k}
-
\eta
\nabla_{\boldsymbol{\theta}}
\mathcal{L}
\left(
\boldsymbol{\theta}^{k}
\right),
\label{eq:update}
\end{equation}
where $\eta$ is the learning rate and $k$ denotes the optimization iteration.
The optimization terminates when the loss function or the inferred parameters have converged.
The present framework does not derive or solve an explicit continuous or discrete adjoint equation. Instead, gradients are obtained by automatic differentiation of the fully discretized computational graph, which propagates sensitivities through every operation of the numerical solver.
Because the loss function is scalar-valued, its gradient with respect to the aerodynamic parameters is computed using reverse-mode automatic differentiation, which efficiently evaluates a sequence of vector--Jacobian products through the differentiable flow solver. 

Beyond parameter inference, the differentiable formulation enables the direct quantification of sensor observability. 
Since the complete flow solver is end-to-end differentiable, the sensitivity of every pressure measurement to each hidden aerodynamic parameter can be evaluated directly through automatic differentiation. For the $i$th sensor and the $j$th hidden parameter, the observability metric is defined as
\begin{equation}
\mathcal{O}_{i}^{(j)}
=
\left[
\frac{1}{N_t}
\sum_{n=1}^{N_t}
\left(
\frac{\partial p_i(t_n)}
{\partial \theta_j}
\right)^2
\right]^{1/2}.
\label{eq:observability}
\end{equation}
Here, $p_i(t_n)$ denotes the pressure measured at the $i$th sensor at time $t_n$, $\theta_j$ is the $j$th component of the hidden aerodynamic parameter vector $\boldsymbol{\theta}$, and $N_t$ is the number of sampled time instants. The observability metric quantifies the root-mean-square sensitivity of the pressure history at a given sensor to perturbations of a particular aerodynamic parameter. 
Large values of $\mathcal{O}_{i}^{(j)}$ indicate that parameter-induced perturbations are strongly reflected in the pressure history recorded by the corresponding sensor, making it highly informative for inverse learning. 
Conversely, small values indicate weak parameter sensitivity and limited observability.

The root-mean-square aggregation provides a compact measure of the cumulative sensitivity of the pressure history over the observation window while remaining independent of the sign of the instantaneous sensitivities. Consequently, sensors exhibiting sustained or repeatedly large parameter sensitivities receive larger observability values than sensors whose responses remain weak throughout the observation interval. The objective of the present study is to rank individual sensor locations according to their overall sensitivity to hidden aerodynamic parameters rather than to characterize the detailed temporal evolution of the sensitivities. Although richer representations that preserve the complete temporal sensitivity history are possible, the scalar observability metric provides a simple, robust, and computationally efficient measure for sensor ranking. Extensions that explicitly incorporate time-resolved sensitivity information constitute an interesting direction for future work.

In the present work, three related concepts are distinguished. Information transport refers to the physical process by which parameter-induced perturbations evolve under the governing compressible Navier--Stokes equations and become encoded in sparse pressure measurements. Observability quantifies the sensitivity of these measurements to hidden aerodynamic parameters through the automatic-differentiation-based metric defined in Eq.~(\ref{eq:observability}). Learnability denotes the practical ability of the gradient-based inverse optimization to accurately recover hidden aerodynamic parameters from the available pressure measurements. Although closely related, these concepts are not equivalent. Information transport provides the physical mechanism through which parameter-induced perturbations reach the sensors, observability quantifies the corresponding parameter-to-measurement sensitivity, and learnability characterizes the effectiveness of the inverse optimization. Consequently, high observability generally facilitates inverse learning but does not necessarily guarantee successful parameter recovery.

The optimization employs reverse-mode automatic differentiation because the loss function is scalar-valued, whereas the observability metric is evaluated using forward-mode automatic differentiation to efficiently compute the parameter sensitivities of the pressure measurements.

The proposed observability metric is a local sensitivity measure evaluated at a prescribed reference parameter state. It quantifies how infinitesimal perturbations of a hidden aerodynamic parameter influence the pressure measurements in the neighborhood of that state. By contrast, inverse learning seeks to minimize the nonlinear loss function from an initial parameter estimate and therefore depends on the global optimization landscape, including its nonlinearity and curvature. Consequently, although large observability generally indicates that a sensor carries strong local information about a parameter, it does not necessarily guarantee rapid or successful convergence during inverse learning because the optimization process is governed by the global loss landscape rather than local sensitivity alone. 
This distinction motivates the subsequent comparison between observability and learnability.

Although the inverse formulation is presented for a general hidden aerodynamic parameter vector, the present study considers only one unknown aerodynamic parameter in each inverse problem. This choice isolates the relationship between sensor observability and inverse learning without the additional complexity introduced by cross-parameter coupling.
Under this setting, the proposed observability metric may be interpreted as the square root of a local time-integrated sensitivity measure and is closely related to the square root of a $1 \times 1$ sensitivity Gramian or Fisher information measure under standard assumptions of independent measurements with uniform noise variance. 
The objective here is to quantify the spatial distribution of information associated with individual sensors for representative single-parameter inverse problems. Extension to simultaneous inference of multiple unknown aerodynamic parameters would require the full sensitivity Gramian or Fisher information matrix, including cross-parameter terms, to assess parameter identifiability, and is left for future work.

For visualization and comparison, the observability values are normalized by the maximum value over all candidate sensor locations,
\begin{equation}
\hat{\mathcal{O}}_{i}^{(j)}
=
\frac{\mathcal{O}_{i}^{(j)}}
{\max_i \mathcal{O}_{i}^{(j)}}.
\end{equation}
The normalized observability map highlights the relative spatial distribution of parameter observability and facilitates the identification of informative sensor locations for inverse learning.

The observability metric provides a quantitative measure of how information about hidden aerodynamic parameters is transported through the flow and encoded in sparse pressure measurements. 
Regions with large observability values exhibit greater sensitivity to the corresponding parameter and are therefore expected to contribute more strongly to parameter identifiability and inverse learning.

The present results suggest that aerodynamic information is transported primarily through the flow field generated by the pitching motion. Variations in the pitching amplitude modify the instantaneous surface-pressure distribution, producing parameter-induced pressure perturbations that propagate through the compressible flow. Compression and expansion waves redistribute these perturbations, while shear layers and wake structures convect them downstream. Consequently, the spatial distribution of observability reflects how these perturbations are transported and redistributed throughout the flow before being encoded in sparse pressure measurements. Regions exhibiting large observability therefore correspond to locations where the parameter-induced perturbations produce the strongest measurable response, providing a physical basis for observability-guided sensor placement.

The proposed framework unifies forward simulation, inverse parameter estimation, and observability analysis within a single differentiable compressible Navier--Stokes formulation. More fundamentally, it establishes a quantitative connection between hidden aerodynamic parameters and their observable signatures in sparse pressure measurements. The governing equations determine how parameter-induced perturbations are generated, transported, and ultimately encoded in sparse measurements, while automatic differentiation reveals these information pathways through end-to-end sensitivity analysis. Within this framework, parameter inference, observability, and sensor placement are not independent problems, but different manifestations of the same underlying process of information transport in compressible aerodynamics.

Table~\ref{tab:parameters} summarizes the baseline numerical parameters used throughout the present study unless otherwise stated. The forward simulations are performed on a computational domain of $(x/c,y/c)\in[0,2]\times[0,6]$ using a uniform Cartesian grid of $1200\times400$ cells, where $c$ denotes the airfoil chord length. To reduce the computational cost of gradient-based optimization, the inverse-learning calculations are performed on a reduced computational domain of $(x/c,y/c)\in[0,2]\times[0,4]$ with a grid resolution of $100\times50$. Both the pitching and heaving phases are prescribed to be zero unless otherwise stated.

\begin{table}[t]
\centering
\caption{Summary of the flow configuration and simulation parameters used in the present study.}
\label{tab:parameters}
\begin{tabular}{llll}
\hline
\multicolumn{2}{c}{Flow parameters} &
\multicolumn{2}{c}{Motion parameters} \\
\hline
Airfoil & NACA 0012, 4412, 6412 &
Pitch amplitude, $A_p$ & $20^\circ$ \\
Mach number, $M_\infty$ & 2.0 &
Heave amplitude, $A_h$ & $0.05c$ \\
Angle of attack, $\alpha$ & $0^\circ$, $2^\circ$, $10^\circ$, $14^\circ$ &
Pitch frequency, $f_p$ & 1.5 Hz \\
CFL number & 0.05 &
Heave frequency, $f_h$ & 0.5 Hz \\
\hline
\end{tabular}
\end{table}

\section{Information Transport and Observability} 
\label{sec:results}

\begin{figure}
\begin{center}
\includegraphics [width=1.\columnwidth]{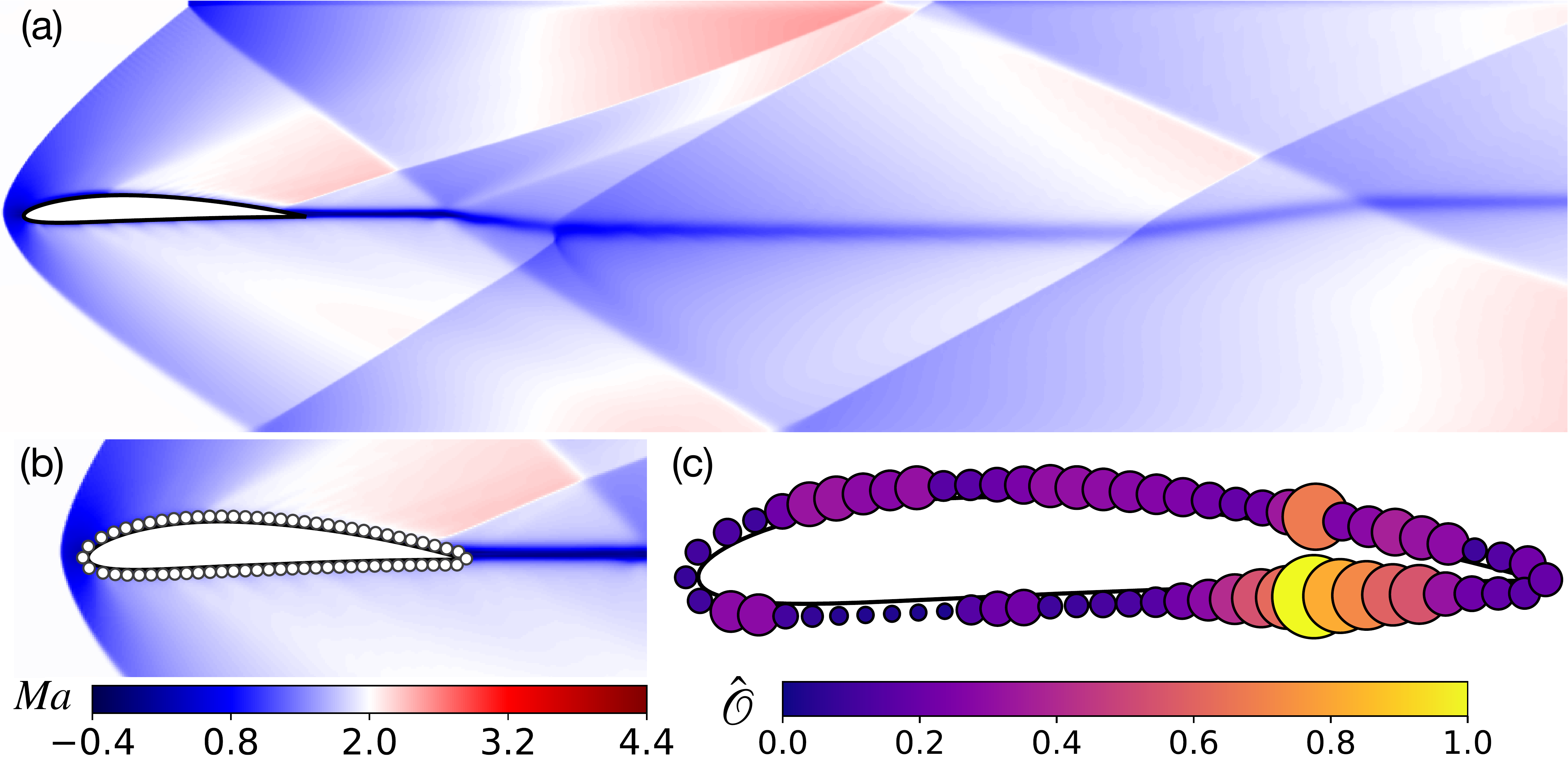}
\end{center}
\caption{Observability distribution for pitch-amplitude inference. (a) Mach-number field around a NACA 4412 airfoil at $M_{\infty}=2.0$ and $\alpha=0^\circ$. (b) Distribution of the 66 pressure probes. (c) AD-based observability map for the pitching amplitude $A_p$. 
}
\label{fig:observability}
\end{figure}  

Figure \ref{fig:observability} illustrates how aerodynamic information associated with the pitching amplitude is transported through the compressible flow and encoded in sparse pressure measurements. Figure \ref{fig:observability}(a) presents the instantaneous Mach-number field obtained from a high-resolution simulation around a NACA 4412 airfoil at $M_{\infty}=2.0$ and $\alpha=0^\circ$, illustrating the principal compressible flow structures. Figure \ref{fig:observability}(b) shows the distribution of 66 virtual pressure probes positioned at a constant normal offset from the airfoil surface.

The AD-based observability metric is evaluated with respect to the pitching amplitude $A_p$, and the resulting distribution is shown in figure \ref{fig:observability}(c). The observability is highly non-uniform, with the largest values concentrated along the lower aft portion of the airfoil and substantially smaller values near the leading edge and portions of the upper surface. This distribution demonstrates that information associated with the pitching motion is redistributed non-uniformly by the compressible flow, producing localized sensing regions with enhanced parameter sensitivity. The observed observability pattern is physically consistent with the compressible flow structures illustrated by the high-resolution simulation in figure \ref{fig:observability}(a). Variations in the pitching amplitude modify the instantaneous surface-pressure distribution and generate pressure perturbations that propagate through the flow. Compression and expansion waves, shear layers, and downstream wake structures redistribute these perturbations and determine how information about the hidden pitching amplitude is conveyed to the sparse pressure measurements. Consequently, the measured pressure response becomes most sensitive in regions where the cumulative effects of pitching-induced pressure modulation and downstream convection are strongest. For the present flow condition, this produces enhanced observability near the lower aft portion of the airfoil, whereas the leading-edge region and much of the upper surface exhibit comparatively weaker sensitivity to the pitching amplitude. The observability map therefore provides a quantitative prediction of the most informative sensor locations, whose effectiveness for parameter inference is examined in the following inverse-learning experiments.

\begin{figure}
\begin{center}
\includegraphics [width=1.\columnwidth]{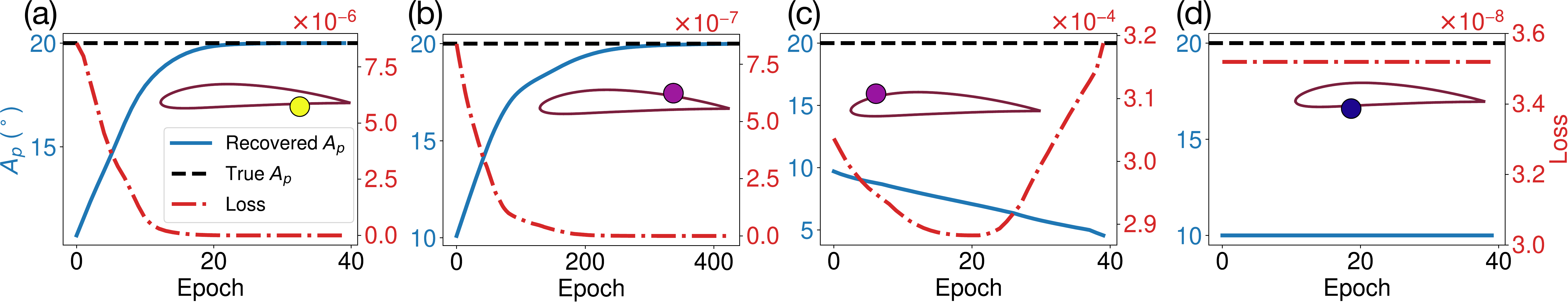}
\end{center}
\caption{Influence of sensor location on inverse learning of the pitching amplitude. Panels (a)-(d) correspond to representative probes with high ($\hat{\mathcal{O}}=1$), moderate-successful ($\hat{\mathcal{O}}=0.336$), moderate-failed ($\hat{\mathcal{O}}=0.322$), and lowest observability ($\hat{\mathcal{O}}=0.024$), respectively.  
}
\label{fig:pitch_loss}
\end{figure}

The observability map predicts the relative informativeness of individual sensors. To examine its relationship with inverse learnability, representative probes spanning the full range of observability values are selected and used individually to recover the pitching amplitude. Figure~\ref{fig:pitch_loss}(a) shows that the highest-observability probe rapidly converges to the true pitching amplitude, whereas the lowest-observability probe in figure~\ref{fig:pitch_loss}(d) exhibits negligible parameter update throughout the optimization. More importantly, figures~\ref{fig:pitch_loss}(b,c) compare two probes with nearly identical observability values but markedly different learning behavior. Although both probes possess comparable local sensitivity to the pitching amplitude, only one successfully recovers the true parameter, while the other fails to converge. These results demonstrate that observability provides an important indicator of inverse learnability but is not sufficient to uniquely determine optimization performance. Successful parameter recovery depends not only on the local parameter-to-measurement sensitivity quantified by the observability metric but also on the evolution of the optimization over the global loss landscape. Consequently, sensors with similar observability may exhibit substantially different convergence histories because inverse learning is governed by global optimization dynamics rather than local sensitivity alone.

\begin{figure}
\begin{center}
\includegraphics [width=1.\columnwidth]{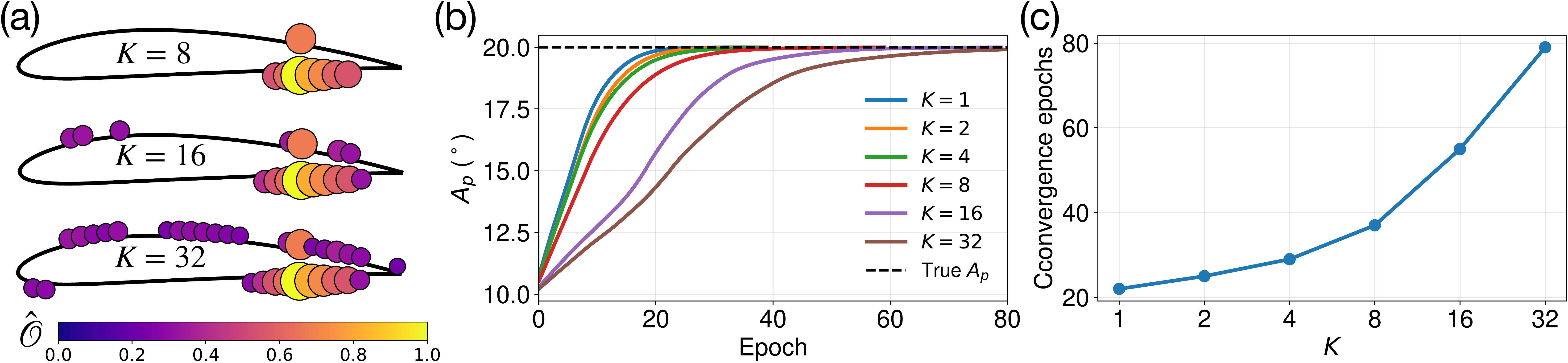}
\end{center}
\caption{Influence of sensor count on inverse learning of the pitching amplitude. (a) Top-$K$ probes selected according to the observability ranking. (b) Recovery histories obtained using the Top-$K$ probe sets. (c) Number of optimization epochs required to satisfy $|A_p-A_{p,\mathrm{true}}|<0.1^\circ$.  
}
\label{fig:topK_sensor_count}
\end{figure}

The single-probe experiments demonstrate that sensor location strongly influences inverse-learning performance. A natural question is whether increasing the number of sensors necessarily improves parameter inference. 
To address this issue, the highest-ranked probes identified by the observability analysis are progressively incorporated into the measurement set.

Figure \ref{fig:topK_sensor_count}(a) illustrates the Top-$K$ sensor sets selected from the observability ranking. The corresponding inverse-learning histories are presented in figure \ref{fig:topK_sensor_count}(b). All Top-$K$ sensor sets successfully recover the true pitching amplitude, indicating that the dominant aerodynamic information is captured by the highest-ranked probes. 

It should be noted that the Top-$K$ strategy ranks sensors solely according to their individual observability values and does not explicitly account for correlations among measurements. Consequently, neighboring sensors located within the same region of high observability may provide redundant information. The objective of the present study is not to develop an optimal multi-sensor placement algorithm but rather to investigate how local sensor observability influences inverse learnability. More sophisticated sensor-selection strategies that explicitly account for measurement redundancy, such as Fisher-information maximization, D-optimal experimental design, greedy information-gain methods, or clustering-based approaches, could be naturally combined with the proposed observability framework and will be investigated in future work.

Surprisingly, however, the convergence rate decreases as additional sensors are incorporated. This trend is quantified in figure~\ref{fig:topK_sensor_count}(c), which shows the number of optimization epochs required to satisfy $|A_p-A_{p,\mathrm{true}}|<0.1^\circ$. The convergence time increases monotonically with sensor count, demonstrating that additional measurements do not necessarily improve inference efficiency. Instead, a small number of highly informative sensors provides the most efficient parameter recovery. This behavior suggests that incorporating additional measurements with lower observability may introduce redundant information and increase the complexity of the optimization, thereby slowing gradient-based convergence.

\begin{figure}
\begin{center}
\includegraphics [width=1.\columnwidth]{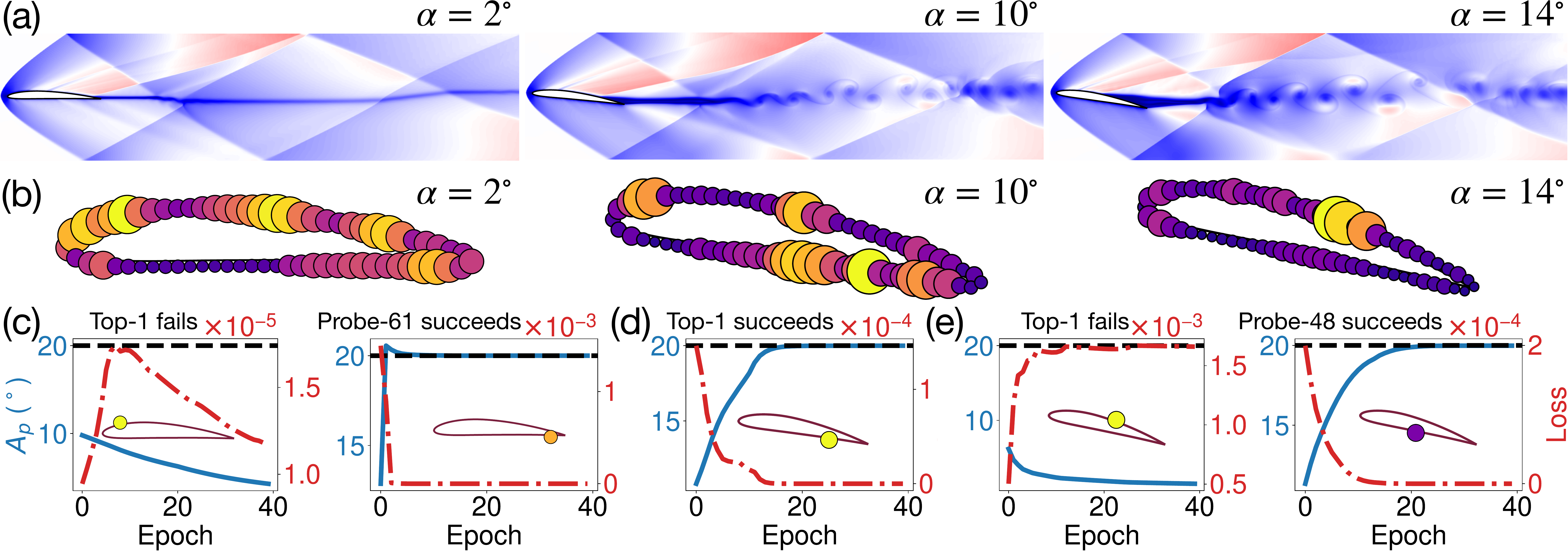}
\end{center}
\caption{Effect of angle of attack on probe learnability. (a) Mach-number fields at $\alpha=2^\circ$, $10^\circ$, and $14^\circ$. (b) Corresponding AD-based observability distributions for the pitching amplitude $A_p$. (c--e) Single-probe inverse learning of $A_p$ at $\alpha=2^\circ$, $10^\circ$, and $14^\circ$, respectively.   
}
\label{fig:flow_aoa}
\end{figure}

Figure~\ref{fig:flow_aoa} investigates the influence of the flow regime on probe learnability at $M_\infty=2$. Three representative angles of attack, $\alpha=2^\circ$, $10^\circ$, and $14^\circ$, are considered. As shown in figure~\ref{fig:flow_aoa}(a), increasing the angle of attack substantially modifies the Mach-number field around the airfoil, altering the compression and expansion waves, shear layers, and downstream wake structures. These changes redistribute the pitching-induced pressure perturbations throughout the flow and consequently modify the corresponding observability distributions, as shown in figure~\ref{fig:flow_aoa}(b). The spatial distribution of information associated with the pitching amplitude is therefore strongly dependent on the underlying flow regime, leading to distinct sensor observability patterns and, consequently, different inverse-learning performance.

The corresponding inverse-learning results are presented in figures~\ref{fig:flow_aoa}(c--e). At $\alpha=10^\circ$, the highest-observability probe rapidly and accurately recovers the pitching amplitude, as shown in figure~\ref{fig:flow_aoa}(d). In contrast, for both $\alpha=2^\circ$ and $14^\circ$, the highest-ranked probe fails to recover $A_p$, whereas a lower-ranked probe converges successfully, as shown in figures~\ref{fig:flow_aoa}(c) and \ref{fig:flow_aoa}(e). These results demonstrate that the sensor with the largest observability is not necessarily the most learnable. Although the observability metric quantifies the local parameter-to-measurement sensitivity, successful inverse learning is also governed by the global optimization behavior, which varies with the underlying flow regime. Consequently, changes in the flow physics can alter the relationship between observability and learnability, indicating that sensor selection based solely on observability magnitude does not always yield the most reliable parameter inference.

\begin{figure}
\begin{center}
\includegraphics [width=1.\columnwidth]{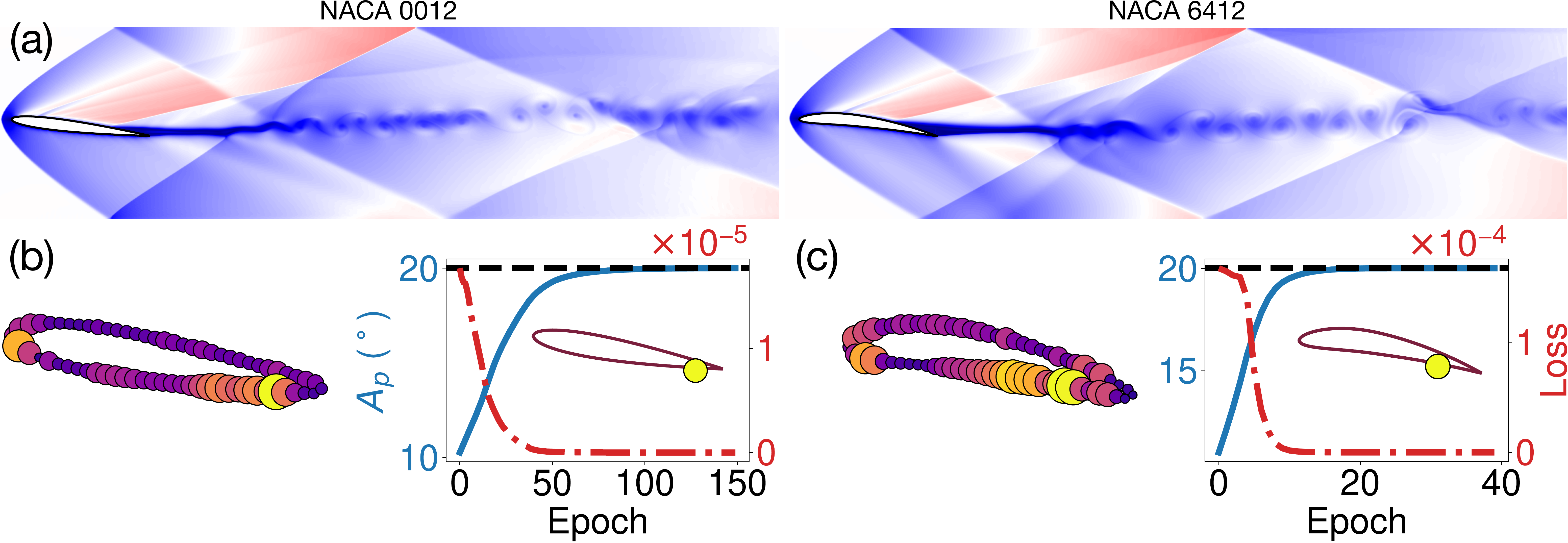}
\end{center}
\caption{Generalization to different airfoil geometries. (a) Mach-number fields around the NACA 0012 and NACA 6412 airfoils at $M_\infty=2$ and $\alpha=10^\circ$. (b,c) AD-based observability distributions for the pitching amplitude $A_p$ together with the corresponding single-probe inverse-learning histories for the NACA 0012 and NACA 6412 airfoils, respectively.   
}
\label{fig:compare_airfoil}
\end{figure}

To assess the generality of the proposed framework, two additional airfoil geometries, namely the symmetric NACA 0012 and the highly cambered NACA 6412 airfoils, are considered under the same operating condition ($M_\infty=2$ and $\alpha=10^\circ$). Owing to their different geometries, both airfoils produce distinct compressible flow fields compared with the NACA 4412 baseline, as shown in figure~\ref{fig:compare_airfoil}(a). These differences redistribute the pitching-induced pressure perturbations and consequently modify the corresponding observability distributions. As shown in figure~\ref{fig:compare_airfoil}(b), the NACA 0012 airfoil exhibits a substantially different observability pattern from that of the NACA 4412 baseline, while its highest-observability probe still accurately recovers the pitching amplitude, albeit with a slower convergence rate. Similarly, figure~\ref{fig:compare_airfoil}(c) shows that the NACA 6412 airfoil possesses a markedly different observability distribution, yet its highest-ranked probe rapidly converges to the true pitching amplitude. These results demonstrate that airfoil geometry significantly influences the compressible flow structures, the resulting observability distribution, and the convergence characteristics of inverse learning. Nevertheless, the proposed AD-based framework consistently identifies informative sensing locations and accurately recovers the pitching amplitude across airfoils with substantially different geometric characteristics.

The present results further suggest that the discrepancy between observability and learnability arises from the distinction between local parameter-to-measurement sensitivity and global optimization behavior. The proposed observability metric quantifies the local sensitivity of the pressure measurements to the hidden aerodynamic parameter at a prescribed reference state, whereas inverse learning depends on the evolution of the optimization throughout the parameter space. Consequently, sensors with the largest local observability do not necessarily exhibit the fastest or most reliable parameter recovery, while sensors with slightly lower observability may converge more rapidly or robustly. The results obtained for different flow regimes and airfoil geometries further indicate that both the spatial distribution of local sensitivities and the optimization behavior are governed by the underlying compressible flow physics. These findings therefore demonstrate that observability provides an important but not sufficient indicator of inverse learnability, highlighting the need to distinguish local parameter sensitivity from global optimization performance when designing sparse sensing strategies for inverse aerodynamic problems.

Taken together, these results demonstrate that automatic differentiation provides not only an efficient optimization framework for aerodynamic inverse problems but also a quantitative tool for analyzing how aerodynamic information is transported and encoded in compressible flows. The proposed observability metric successfully identifies informative sensing regions and provides a physics-informed criterion for sensor placement, while the inverse-learning experiments demonstrate that observability alone does not uniquely determine learnability. Instead, the relationship between observability and learnability is governed by the underlying compressible flow physics, with both the flow regime and the airfoil geometry substantially influencing the spatial distribution of observability and the convergence behavior of parameter inference. Nevertheless, across all cases considered, the proposed framework consistently identifies effective sensing locations and accurately recovers the pitching amplitude, demonstrating its robustness and generality for sparse aerodynamic sensing and inverse learning.

\section{Conclusion}
\label{sec:conclusion}

An automatic-differentiation-based framework has been developed to investigate information transport and aerodynamic parameter inference in compressible flows. By differentiating a fully differentiable immersed-boundary flow solver, an observability metric was introduced to quantify the sensitivity of sparse pressure measurements to hidden aerodynamic parameters and to guide sensor placement for inverse learning.

The proposed framework demonstrates that aerodynamic information is transported non-uniformly through the compressible flow, producing localized sensing regions with enhanced parameter sensitivity. The inverse-learning experiments further reveal that observability and learnability are closely related but fundamentally distinct concepts. Although highly observable probes generally provide more effective parameter recovery, the highest-observability probe is not consistently the most learnable under the operating conditions investigated. Moreover, both the flow regime and the airfoil geometry substantially influence the spatial distribution of observability and the convergence behavior of inverse learning, highlighting the important role of the underlying flow physics in governing the relationship between information transport, observability, and learnability.

Although the differentiable framework is formulated for a general aerodynamic parameter vector, the present study focuses on representative single-parameter inverse problems to isolate the relationship between sensor observability and inverse learning. The formulation is readily extendable to additional aerodynamic parameters, while simultaneous multi-parameter inference, cross-parameter identifiability, and robustness under measurement uncertainty remain important directions for future investigation. In addition, the present Top-$K$ sensor ranking is based on individual sensor observability and does not explicitly account for redundancy among multiple measurements. Extending the proposed framework to jointly optimize multiple sensor locations using Fisher-information maximization, D-optimal experimental design, or related information-theoretic sensor placement strategies represents another promising direction for future work.

Taken together, these results demonstrate that automatic differentiation can serve not only as an optimization technique but also as a scientific tool for investigating how aerodynamic information is transported and encoded in compressible flows. The proposed framework establishes a quantitative connection between information transport, sensor observability, and inverse learnability, providing a physics-based foundation for observability-guided sensor placement and aerodynamic inverse analysis. More broadly, the framework offers a foundation for future differentiable methodologies for sparse sensing, aerodynamic system identification, and simultaneous multi-parameter inference in compressible flows.

\section*{Acknowledgements}
This research was supported by the Division of Research and Innovation Partnership Commitments (RIPS) at Northern Illinois University. 

\section*{Data availability}
The data that support the findings of the present work are available from the corresponding author upon reasonable request. 

\section*{References}
\bibliographystyle{unsrtnat}
\bibliography{refs}

@article{Fukami_2025b,
  title={Observable-augmented manifold learning for multi-source turbulent flow data},
  author={Fukami, Kai and Taira, Kunihiko},
  journal={J.~Fluid Mech.},
  volume={1010},
  pages={R4},
  year={2025},
  publisher={Cambridge University Press}
}

@article{Zhang_2026a,
  title={A differentiable, shock-capturing neural solver for compressible flow simulation},
  author={Zhang, Bo},
  journal={Phys.~Fluids},
  volume = {38},
  number = {4},
  pages = {046105},
  year = {2026},
  month = {04},
  publisher={AIP Publishing}
}

@article{Da_2020a,
  title={Flow state estimation in the presence of discretization errors},
  author={da Silva, Andre FC and Colonius, Tim},
  journal={J.~Fluid Mech.},
  volume={890},
  pages={A10},
  year={2020},
  publisher={Cambridge University Press}
}

@article{Fukami_2025a,
  title={Compact representation of transonic airfoil buffet flows with observable-augmented machine learning},
  author={Fukami, Kai and Iwatani, Yuta and Maejima, Soju and Asada, Hiroyuki and Kawai, Soshi},
  journal={J.~Fluid Mech.},
  volume={1021},
  pages={A39},
  year={2025},
  publisher={Cambridge University Press}
}

@article{Ma_2020a,
  title={Time-resolved topology of turbulent boundary layer separation over the trailing edge of an airfoil},
  author={Ma, Austin and Gibeau, Bradley and Ghaemi, Sina},
  journal={J.~Fluid Mech.},
  volume={891},
  pages={A1},
  year={2020},
  publisher={Cambridge University Press}
}

@article{Li_2018a,
  title={Vortex force map method for viscous flows of general airfoils},
  author={Li, Juan and Wu, Zi-Niu},
  journal={J.~Fluid Mech.},
  volume={836},
  pages={145--166},
  year={2018},
  publisher={Cambridge University Press}
}

@article{Hartmann_2013a,
  title={Coupled airfoil heave/pitch oscillations at buffet flow},
  author={Hartmann, Axel and Klaas, Michael and Schr{\"o}der, Wolfgang},
  journal={AIAA~J.},
  volume={51},
  number={7},
  pages={1542--1552},
  year={2013},
  publisher={American Institute of Aeronautics and Astronautics}
}

@article{Himpe_2018a,
  title={emgr--The empirical gramian framework},
  author={Himpe, Christian},
  journal={Algorithms},
  volume={11},
  number={7},
  pages={91},
  year={2018},
  publisher={MDPI}
}

@article{Wang_2005a,
  title={Information-theoretic approaches for sensor selection and placement in sensor networks for target localization and tracking},
  author={Wang, Hanbiao and Yao, Kung and Estrin, Deborah},
  journal={J.~Commun.~Netw.},
  volume={7},
  number={4},
  pages={438--449},
  year={2005},
  publisher={KICS}
}

@article{Loiseau_2018a,
  title={Sparse reduced-order modelling: sensor-based dynamics to full-state estimation},
  author={Loiseau, Jean-Christophe and Noack, Bernd R and Brunton, Steven L},
  journal={J.~Fluid Mech.},
  volume={844},
  pages={459--490},
  year={2018},
  publisher={Cambridge University Press}
}

@article{Hinson_2014a,
  title={Observability-based optimal sensor placement for flapping airfoil wake estimation},
  author={Hinson, Brian T and Morgansen, Kristi A},
  journal={J.~Guid.~Control~Dyn.},
  volume={37},
  number={5},
  pages={1477--1486},
  year={2014},
  publisher={American Institute of Aeronautics and Astronautics}
}

@inproceedings{Ke_2026a,
  title={Observability of Initial States in One-Dimensional Inviscid Flows With Shocks},
  author={Ke, Guanguan and Grauer, Samuel J and Zaki, Tamer A},
  booktitle={AIAA SCITECH 2026 Forum},
  pages={0493},
  year={2026}
}

@article{Buchta_2022a,
  title={Assimilation of wall-pressure measurements in high-speed flow over a cone},
  author={Buchta, David A and Laurence, Stuart J and Zaki, Tamer A},
  journal={J.~Fluid Mech.},
  volume={947},
  pages={R2},
  year={2022},
  publisher={Cambridge University Press}
}

@article{Wang_2022b,
  title={What is observable from wall data in turbulent channel flow?},
  author={Wang, Qi and Wang, Mengze and Zaki, Tamer A},
  journal={J.~Fluid Mech.},
  volume={941},
  pages={A48},
  year={2022},
  publisher={Cambridge University Press}
}

@article{Wang_2021b,
  title={State estimation in turbulent channel flow from limited observations},
  author={Wang, Mengze and Zaki, Tamer A},
  journal={J.~Fluid Mech.},
  volume={917},
  pages={A9},
  year={2021},
  publisher={Cambridge University Press}
}

@article{Brunton_2016a,
  title={Discovering governing equations from data by sparse identification of nonlinear dynamical systems},
  author={Brunton, Steven L and Proctor, Joshua L and Kutz, J Nathan},
  journal={Proc.~Natl.~Acad.~Sci.~U.S.A.},
  volume={113},
  number={15},
  pages={3932--3937},
  year={2016},
  publisher={National Acad Sciences}
}

@article{Mittal_2005a,
	Author = {Mittal, R. and Iaccarino, G.},
	Journal = {Annu.~Rev.~Fluid Mech.},
	Pages = {239-261},
	Title = {Immersed boundary methods},
	Volume = {37},
	Year = {2005}}

\end{document}